\documentclass{ws-procs9x6}

\def\trimmarks{}
\def\draftnote{\hfill}

\begin{document}

\title{Supernova Neutrinos:\\ Flavor-Dependent Fluxes and Spectra}

\author{GEORG G.~RAFFELT and MATHIAS Th.~KEIL}

\address{Max-Planck-Institut f\"ur Physik (Werner-Heisenberg-Institut)\\
F\"ohringer Ring 6, 80805 M\"unchen, Germany}

\author{ROBERT BURAS, HANS-THOMAS JANKA and MARKUS RAMPP}

\address{Max-Planck-Institut f\"ur Astrophysik\\
Karl-Schwarzschild-Str.~1, 85741 Garching, Germany}

\maketitle

\vskip-18pt
\vbox to 0pt{{\ }\newline\vskip-9cm \noindent
\normalsize\rm Contribution to Proc.\ 
{\it NOON 2003: The 4th Workshop on Neutrino Oscillations 
and their Origin},
10--14 February 2003, Kanazawa, Japan.\vfil}

\abstracts{Transporting $\nu_\mu$ and $\nu_\tau$ in a supernova (SN)
  core involves several processes that have been neglected in
  traditional simulations.  Based on a Monte Carlo study we find that
  the flavor-dependent spectral differences are much smaller than is
  often stated in the literature. A full-scale SN simulation using a
  Boltzmann solver and including all relevant neutrino reactions
  confirms these results.  The flavor-dependent flux differences are
  largest during the initial accretion phase.}

\section{Introduction}

A supernova (SN) core is essentially a blackbody neutrino source, but
in detail the fluxes and spectra depend on the flavor. Up to very
small details $\nu_\mu$, $\nu_\tau$, $\bar\nu_\mu$ and $\bar\nu_\tau$
can be treated on an equal footing and will be collectively refered to
as $\nu_\mu$. Numerical simulations usually find a hierarchy $\langle
E_{\nu_e}\rangle<\langle E_{\bar\nu_e}\rangle <\langle
E_{\nu_\mu}\rangle$ and approximately equal luminosities.  The
spectral differences offer an opportunity to observe flavor
oscillations as the source fluxes will get partially interchanged. For
example, it may be possible to distinguish a normal from an inverted
neutrino mass hierarchy%
\cite{Dighe:1999bi,Lunardini:2003eh,Takahashi:2002cm,Takahashi:2001dc}.

A full-scale numerical simulation by the Livermore group finds for the
integrated signal $\langle E_{\nu_e}\rangle=13$, $\langle
E_{\bar\nu_e}\rangle=16$ and $\langle E_{\nu_\mu}\rangle=23$~MeV and
almost perfect equipartition of the luminosities\cite{Totani:1997vj},
results that are representative for traditional numerical simulations.
Sometimes extreme spectral hierarchies of up to $\langle
E_{\bar\nu_e}\rangle:\langle E_{\nu_\mu}\rangle \approx 1:2$ have been
stated, but searching the literature we find no support for such
claims by credible simulations\cite{Keil:2002in}.

Traditional numerical simulations treat the $\nu_\mu$ and $\nu_\tau$
transport somewhat schematically because their exact fluxes and
spectra may not be important for the explosion mechanism. When a
number of missing reactions are included one finds that $\langle
E_{\bar\nu_e}\rangle$ and $\langle E_{\nu_\mu}\rangle$ are much more
similar than had been thought. The remaining spectral and flux
differences are probably large enough to observe oscillation effects
in a high-statistics galactic SN signal, but the details are more
subtle than had been assumed in the past.

\section{Mu- and Tau-Neutrino Transport}

The transport of $\nu_e$ and $\bar\nu_e$ is dominated by $\nu_e
n\leftrightarrow p e^-$ and $\bar\nu_e p\leftrightarrow n e^+$,
reactions that freeze out at the energy-dependent ``neutrino sphere.''
The flux and spectrum is essentially determined by the temperature and
geometric size of this emission region.  Moreover, the neutron density
is larger than that of protons so that the $\bar\nu_e$ sphere is
deeper than the $\nu_e$ sphere, explaining $\langle
E_{\nu_e}\rangle<\langle E_{\bar\nu_e}\rangle$.

For $\nu_\mu$, in contrast, the flux and spectra formation is a
three-step process. The main opacity source is neutral-current nucleon
scattering $\nu_\mu N\to N \nu_\mu$.  Deep in the star thermal
equilibrium is maintained by nucleon bremsstrahlung $N
N\leftrightarrow N N \nu_\mu \bar\nu_\mu$, pair annihilation $e^-
e^+\leftrightarrow \nu_\mu \bar\nu_\mu$ and $\nu_e
\bar\nu_e\leftrightarrow \nu_\mu \bar\nu_\mu$, and scattering on
electrons $\nu_\mu e^-\to e^- \nu_\mu$.  The freeze-out sphere of the
pair reactions defines the ``number sphere,'' that of the
energy-changing reactions the ``energy sphere,'' and finally that of
nucleon scattering the ``transport sphere'' beyond which neutrinos
stream freely. Between the energy and transport spheres the neutrinos
scatter without being absorbed or emitted and without much energy
exchange, i.e.\ in this ``scattering atmosphere'' they propagate by
diffusion.

One may think that the $\nu_\mu$ spectrum is fixed by the medium
temperature at the energy sphere so that $\langle
E_{\bar\nu_e}\rangle<\langle E_{\nu_\mu}\rangle$ because the
energy-sphere is deeper and hotter than the $\bar\nu_e$ sphere.
However, the scattering atmosphere is more opaque to higher-energy
neutrinos because the cross section scales as $E_\nu^2$, biasing the
escaping flux to lower energies. For typical conditions $\langle
E_{\nu_\mu}\rangle$ of the escaping flux is 50--60\% of the value
characteristic for the temperature at the energy
sphere\cite{Raffelt:ai}.  Therefore, the final $\langle
E_{\bar\nu_e}\rangle:\langle E_{\nu_\mu}\rangle$ ratio is the result
of two large counter-acting effects, the large temperature difference
between the $\nu_\mu$ energy sphere and the $\bar\nu_e$ sphere on the
one hand, and the energy-dependent ``filter effect'' of the scattering
atmosphere on the other.

Until recently all simulations simplified the treatment of $\nu_\mu$
transport in that energy-exchange was not permitted in $\nu
N$-scattering, $e^-e^+$ annihilation was the only pair process, and
$\nu_\mu e$-scattering was the only energy-exchange process. However,
it has been recognized for some time that nucleon recoils are
important for energy
exchange\cite{Keil:2002in,Raffelt:ai,Janka:1995ir}, that nucleon
bremsstrahlung is an important pair process%
\cite{Keil:2002in,Raffelt:ai,Janka:1995ir,Suzuki1991,Suzuki1993,%
Hannestad:1997gc,Thompson:2000gv}, and more recently that $\nu_e
\bar\nu_e\to \nu_\mu \bar\nu_\mu$ is far more important than $e^-
e^+\to\nu_\mu \bar\nu_\mu$ as a $\nu_\mu\bar\nu_\mu$ source
reaction\cite{Keil:2002in,Buras:2002wt}.

\begin{figure}[b]
\centerline{\epsfxsize=0.5\hsize\epsfbox{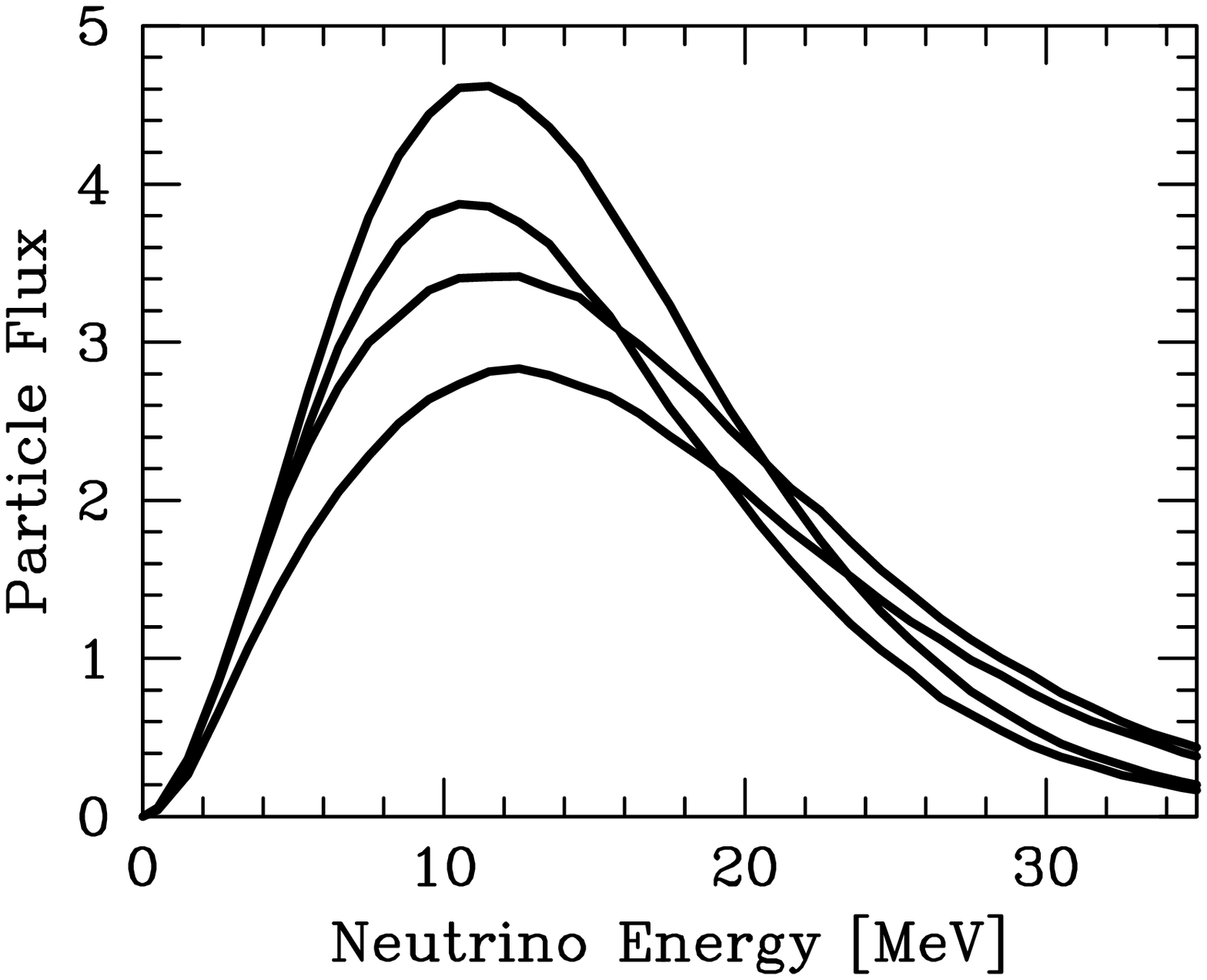}
\epsfxsize=0.5\hsize\epsfbox{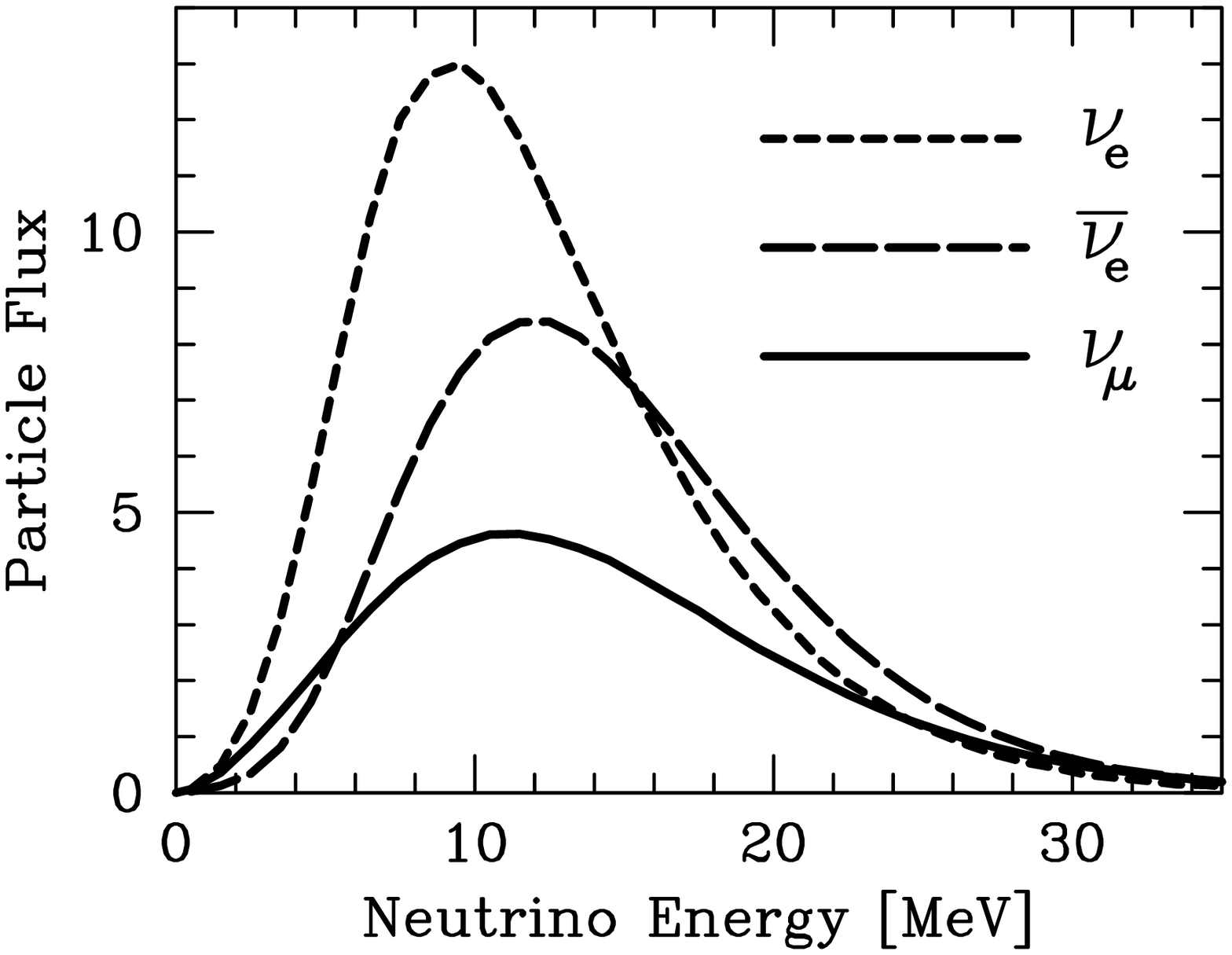}}
\caption{Neutrino fluxes for an accretion-phase model.
{\it Left panel, curves from bottom to top:}~Flux of $\nu_\mu$
with traditional neutrino interaction channels, then
adding nucleon bremsstrahlung, next adding
nucleon recoils, and finally adding $\nu_e\bar\nu_e$ annihilation.
{\it Right~panel:} Fluxes for all flavors; the $\nu_\mu$ curve
includes all reaction channels.
\label{fig:fluxes}}
\end{figure}

We have performed a detailed assessment of the relevance of the new
reactions on the basis of a Monte Carlo study\cite{Keil:2002in}. To
illustrate the results we use a hydrodynamically self-consistent
accretion-phase model and show in Fig.~\ref{fig:fluxes} (left panel)
the $\nu_\mu$ flux spectrum when using the traditional input physics
(bottom curve). Then we add nucleon bremsstrahlung that increases the
flux without changing much the average energy. Next we switch on
nucleon recoils that depletes the spectrum's high-energy tail without
changing much the overall particle flux. Finally, switching on
$\nu_e\bar\nu_e$ annihilation increases the flux without affecting the
spectrum much. The compound effect of the new processes is not overly
dramatic, but so large that all of them should be included in serious
full-scale simulations.

In the right panel of Fig.~\ref{fig:fluxes} we compare for the same
model the flux spectra of $\nu_e$, $\bar\nu_e$ and $\nu_\mu$, the
latter including all reactions. In this example $\langle
E_{\bar\nu_e}\rangle$ almost exactly equals $\langle
E_{\nu_\mu}\rangle$, but the fluxes differ by almost a factor
of~2. This is reverse to the usual assumption of a pronounced
hierarchy of average energies and nearly exact equipartition of the
luminosities.

We have studied a variety of stellar background models, some of them
self-consistent hydrostatic models, others power-law profiles of
density and temperature. For realistic cases we never find extreme
spectral hierarchies, the differences between $\langle
E_{\bar\nu_e}\rangle$ and $\langle E_{\nu_\mu}\rangle$ typically being
0--20\%. On the other hand, the fluxes can be rather different,
especially during the accretion phase when the atmosphere is quite
extended. The different neutrino spheres have then rather different
geometric extensions, explaining large flux differences. Later during
the Kelvin-Helmholtz cooling phase the star is very compact so that
any geometry effect of the radiating surfaces is small. Moreover, the
relevant regions are then neutron rich so that the transport physics
of $\bar\nu_e$ and $\nu_\mu$ will become similar. Therefore, during
the late phases one expects very similar $\bar\nu_e$ and $\nu_\mu$
fluxes, while, of course, the $\nu_e$ flux and spectrum remain
unaffected by our arguments.

\section{Spectral Characteristics}

To characterize the neutrino fluxes one naturally uses some global
parameters such as the particle flux, the luminosity (energy flux),
and the average energy $\langle E\rangle$. In order to characterize
the spectral shape in greater detail one may also invoke higher energy
moments $\langle E^n\rangle$.  One measure frequently given from
numerical simulations is $E_{\rm rms}=\sqrt{\langle E^3\rangle/\langle
E\rangle}$ because of its relevance for calculating average
neutrino-nucleon interaction rates.

Sometimes a global analytic fit to the spectra is also
useful. Frequently one approximates the flux spectra by a nominal
Fermi-Dirac function
\begin{equation}
f(E)\propto \frac{E^2}{1+\exp(-\eta+E/T)}
\end{equation}
with a temperature $T$ and a degeneracy parameter $\eta$.  This
approximation allows one to fit the overall luminosity and two energy
moments, typically chosen to be $\langle E\rangle$ and $\langle
E^2\rangle$. However, the Fermi-Dirac fit is not more natural than
other possibilities; certainly the low- and high-energy tails of the
spectra are not especially well represented by this fit.

We find that the Monte Carlo spectra are approximated over a broader
range of energies by a simpler functional form that we
call ``alpha fit,''
\begin{equation}\label{eq:alpha-fit}
f(E)\propto E^{\alpha} \exp\left[-(\alpha+1)\,{E}/{\bar E}\right]\,.
\end{equation}
For any value of $\alpha$ we have $\langle E\rangle=\bar E$, a
Maxwell-Boltzmann spectrum corresponds to $\alpha=2$.  The numerical
spectra show values of $\alpha=2.5$--5, i.e.\ they are ``pinched.''

\eject

\section{A Full-Scale Simulation}

In the Garching SN code\cite{Rampp:2002} we have now implemented all
relevant neutrino interaction rates, including nucleon bremsstrahlung,
neutrino pair processes, weak magnetism, and nucleon recoils. Our
treatment of neutrino-nucleon interactions includes nuclear
correlation effects.  The transport part of this code is based on a
Boltzmann solver.  The neutrino-radiation hydrodynamics program
enables us to perform spherically symmetric as well as
multi-dimensional simulations, thus allowing us to take into account
the effects of convection.

\begin{figure}[b]
\centerline{\epsfxsize=0.85\hsize\epsfbox{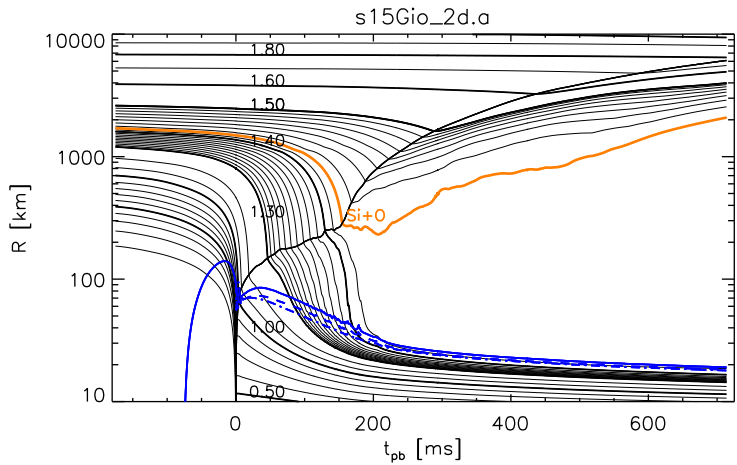}}
\caption{Trajectories of the mass shells in the core of an exploding
15$\,$M$_{\odot}$ star. The explosion occurs about 150$\,$ms after
shock formation, developing a bifurcation (``bubble'') between the
mass that follows the outgoing shock and the mass that settles on the
nascent neutron star. Also indicated are the positions of the neutrino
spheres of $\nu_e$, $\bar\nu_e$ and
$\nu_{\tau}$.\label{fig:explosion}}
\end{figure}

\begin{figure}[b]
\centerline{\epsfxsize=0.5\hsize\epsfbox{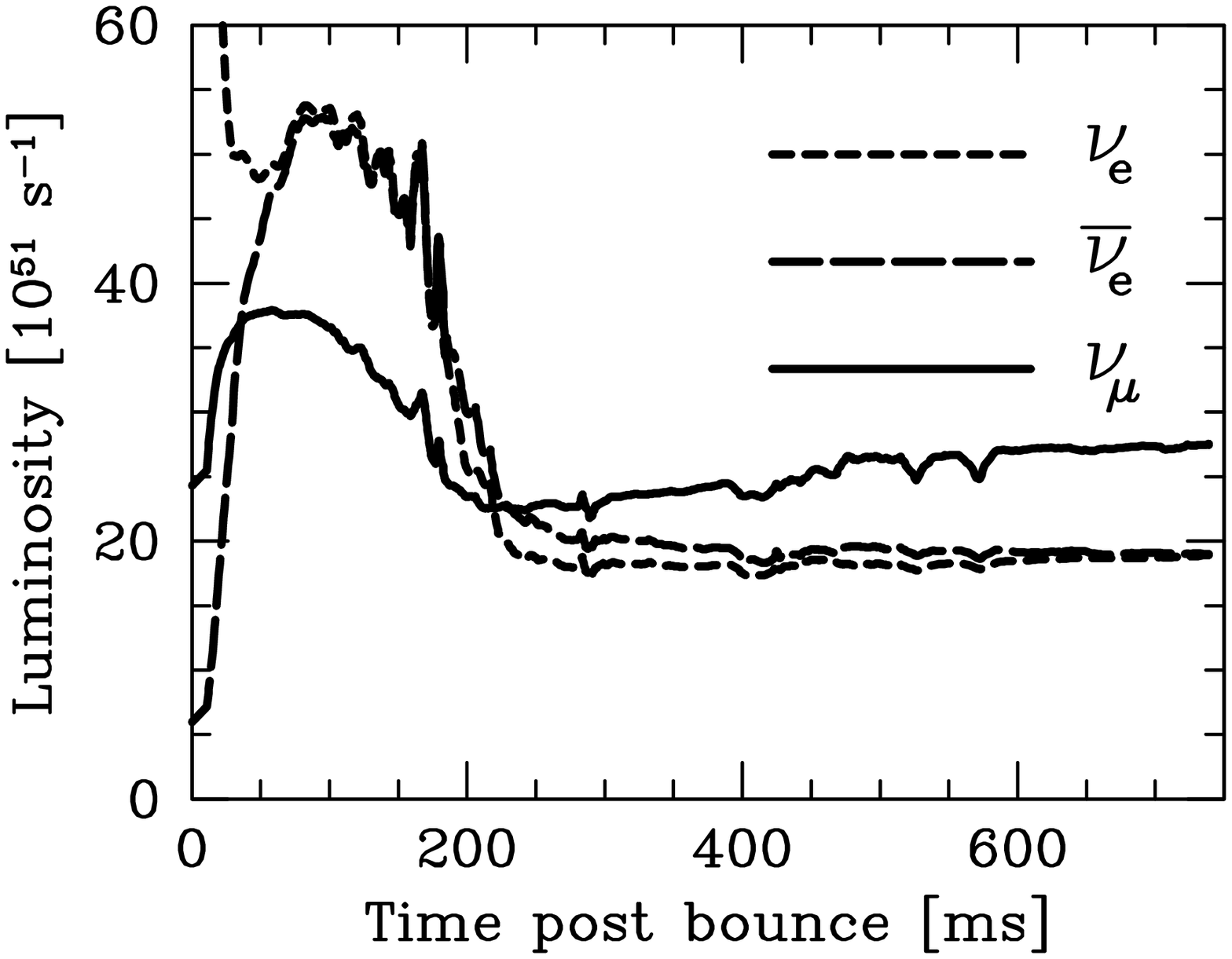}
\epsfxsize=0.5\hsize\epsfbox{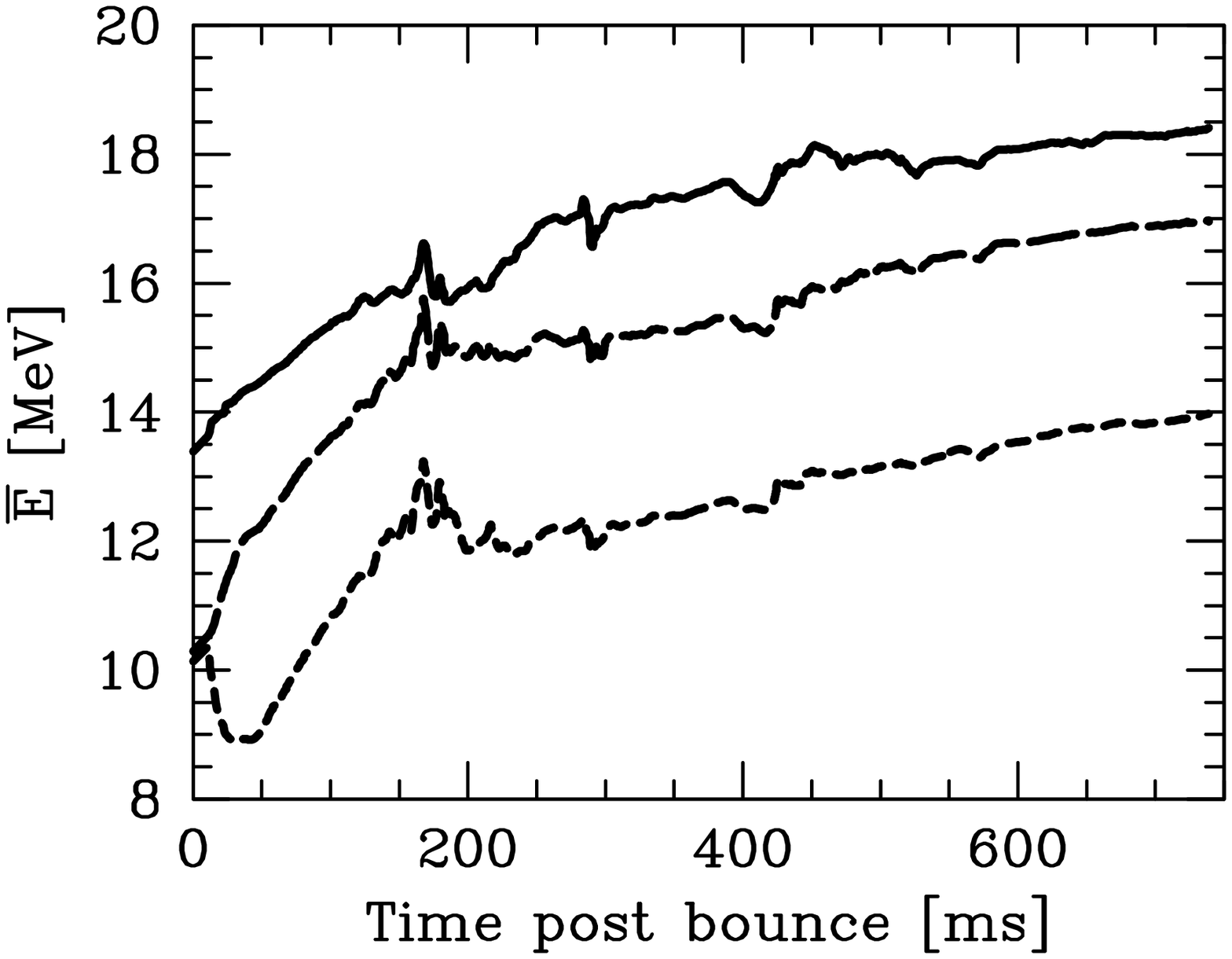}}
\centerline{\epsfxsize=0.5\hsize\epsfbox{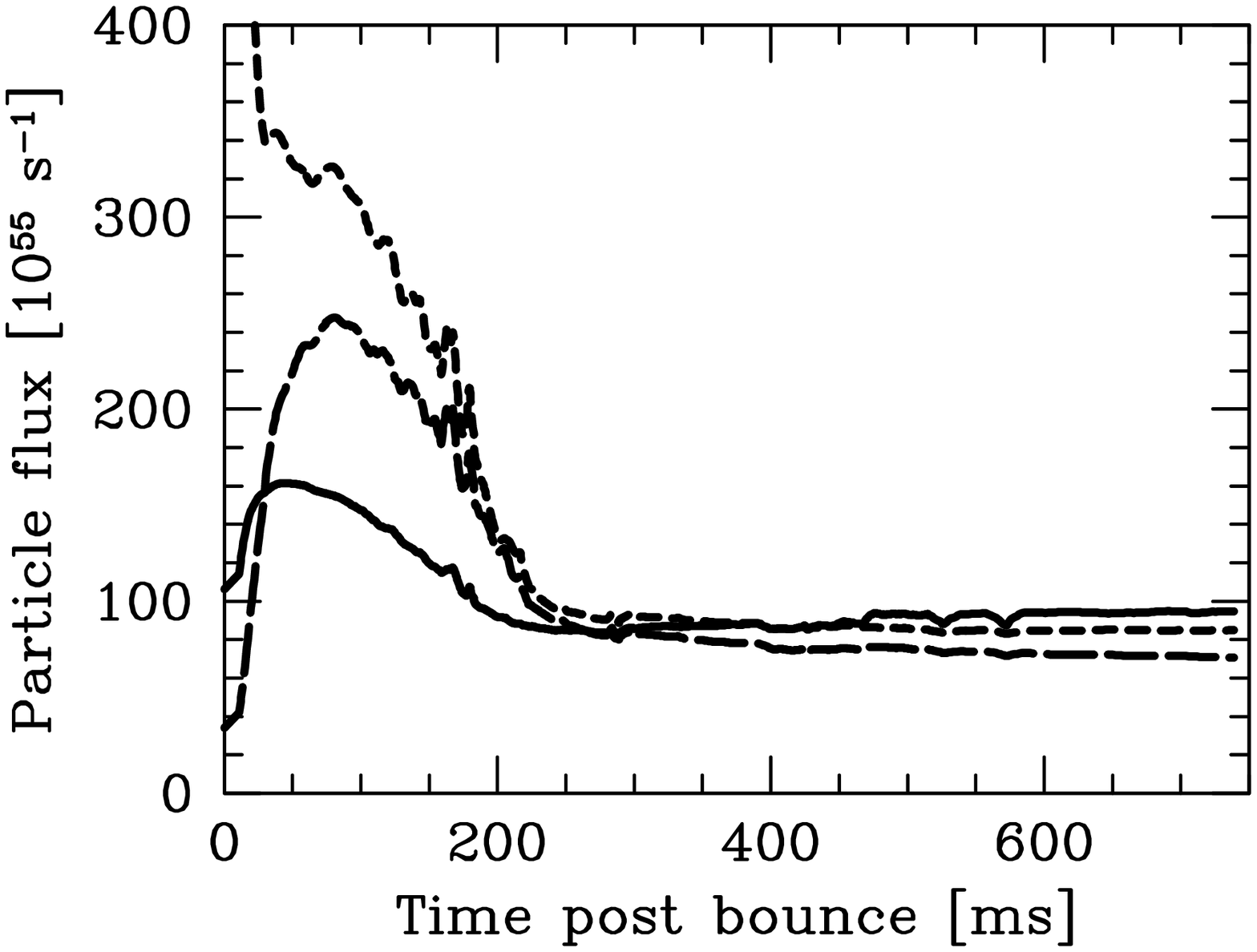}
\epsfxsize=0.5\hsize\epsfbox{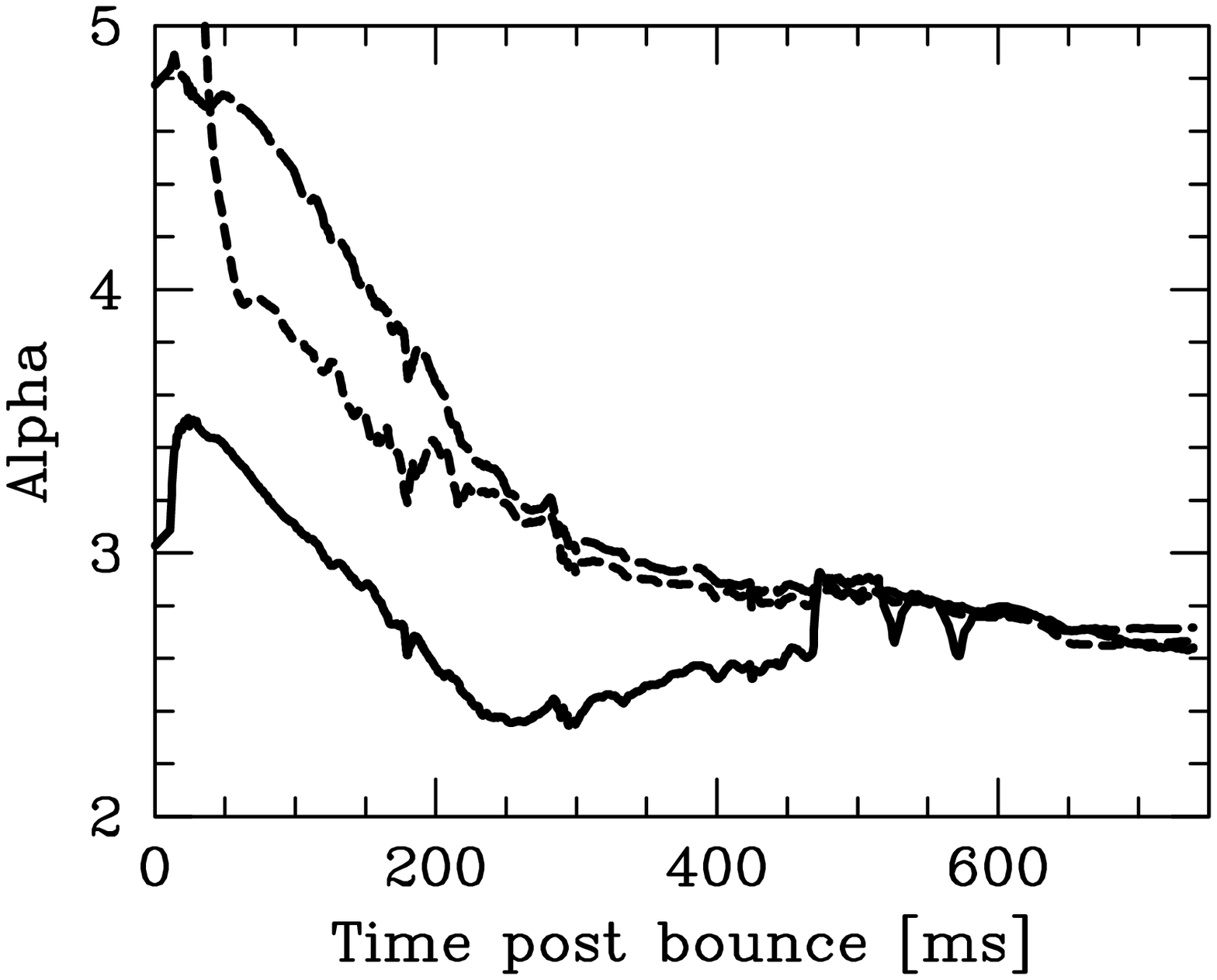}}
\caption{Neutrino fluxes and spectral properties for the full-scale
simulation described in the text. The hydrodynamic bounce and shock
formation occur at $t = 0$ (cf.~Fig.~\ref{fig:explosion}).  The right
upper plot gives the spectral fit parameter $\bar{E}$, the right lower
one $\alpha$. Note that the discontinuity in the latter at $t \approx
480\,$ms is caused by mapping the model from two dimensions to one.
\label{fig:fullscale}}
\end{figure}

To explore the time-dependent properties and long-time evolution of
the neutrino signal, we currently continue a state-of-the-art
hydrodynamic calculation of a SN into the Kelvin-Helmholtz neutrino
cooling phase of the forming neutron star. The progenitor model is a
15$\,$M$_{\odot}$ star with a 1.28$\,$M$_{\odot}$ iron core
(Model~s15s7b2 from S.~Woosley; personal communication).  The period
from shock formation to 480$\,$ms after bounce was evolved in two
dimensions. The subsequent evolution of the model is simulated in
spherical symmetry.  At 150$\,$ms the explosion sets in, driven by
neutrino energy deposition and aided by very strong convective
activity in the neutrino-heating region behind the shock
(Fig.~\ref{fig:explosion}).  Note that a small modification of the
Boltzmann transport was necessary to allow the explosion to
happen\cite{Janka:2002}. Unmanipulated full-scale models with an
accurate treatment of the microphysics currently do not obtain
explosions\cite{Buras:2003}. Details of this run will be documented
elsewhere; at the time of this writing the CPU expensive calculation
is still on the computer.  Here we show in Fig.~\ref{fig:fullscale} a
preview of the main characteristics of the neutrino signal up to
750~ms post bounce.

The neutrino signal agrees with what is expected for the standard
delayed-explosion scenario. In particular, it clearly shows the prompt
$\nu_e$ burst and a broad shoulder in all fluxes during the accretion
phase that ends at 200~ms when the explosion has taken off.  The
average neutrino energies follow the usual hierarchy and they increase
with time due to the contraction of the star.  We also show the alpha
parameter from a global fit according to
Eq.~(\ref{eq:alpha-fit}). During the accretion phase the $\nu_\mu$
flux is least pinched, at late times the $\alpha$ values of all
flavors converge near~2.5.

These results agree with and nicely illustrate our previous Monte
Carlo findings in that the spectral hierarchy between $\bar\nu_e$ and
$\nu_\mu$ is rather mild and in that the average energies converge at
late times.  Conversely, the particle fluxes differ by almost a factor
of 2 during the accretion phase, but cross over shortly after the
explosion.  At 750~ms the differences between the fluxes continue to
increase, an asymptotic value has not yet been reached.

\section{Conclusions}

Traditional numerical SN simulations had two weaknesses regarding the
flavor-dependent neutrino fluxes and spectra. First, the interaction
between $\nu_{\mu,\tau}$ and $\bar\nu_{\mu,\tau}$ and the stellar
medium was schematic, neglecting a number of important
processes. Second, a Boltzmann solver for neutrino transport could not
be coupled self-consistently with the hydrodynamic evolution.

We have performed a systematic Monte Carlo study on various stellar
background models and the first SN simulation that includes all
relevant interaction rates and a Boltzmann solver. While the usual
relationship between the $\nu_e$ and $\bar\nu_e$ fluxes and spectra
remains essentially unchanged, the $\nu_\mu$ spectrum is much more
similar to that of $\bar\nu_e$, especially during the Kelvin-Helmholtz
cooling phase.  Differences of the average energies are in the range
0--20\%, with 10\% being a typical number. During the accretion phase
the $\nu_\mu$ particle flux is smaller than that of $\bar\nu_e$ by up
to a factor of~2, but later the particle fluxes cross over.

Our findings imply that observing neutrino oscillation effects in a SN
signal is a more subtle problem than had been thought previously, but
by no means impossible. However, when exploring the physics potential
of a future galactic SN one should not rely on the notion of an exact
flavor equipartition of the luminosities or the extreme spectral
differences that have sometimes been stated in the literature.

\section*{Acknowledgments}

This work was supported, in part, by the Deutsche
Forschungsgemeinschaft under grant No.\ SFB-375 and by the European
Science Foundation (ESF) under the Network Grant No.~86 Neutrino
Astrophysics.


\end{document}